\documentclass[a4paper,11pt]{article}
\usepackage{jcappub} % for details on the use of the package, please see the JINST-author-manual's
\makeatletter
\gdef\@fpheader{}   % removes "Prepared for submission to JCAP"
\gdef\@journal{}    % often removes extra journal info if any
\makeatother

\usepackage{lineno}
\usepackage{graphicx} 
\usepackage{subcaption}
\usepackage{multirow}
\usepackage{hyperref}
\usepackage[group-separator={,}]{siunitx}
\usepackage{xcolor}

\title{Molecular absorption of Cherenkov light at CTAO}

\author[a]{G. Voutsinas,}
\author[a]{M. Dalchenko,}
\author[b]{M. Gaug,}
\author[c]{O. Gueta,}
\author[a]{T. Montaruli}
\author[d]{and R. Zanin}

\affiliation[a]{Département de Physique Nucléaire et Corpusculaire, Faculté de Sciences, Université de Genève, 1205 Geneva, Switzerland}
\affiliation[b]{Unitat de F\'isica de les Radiacions, Departament de F\'isica, and CERES-IEEC, \\Universitat Aut\`onoma de Barcelona, E-08193 Bellaterra, Spain}
\affiliation[c]{CTAO, Science Data Management Centre (SDMC), Platanenallee 6, 15738 Zeuthen, Germany}
\affiliation[d]{CTAO, Via Piero Gobetti 93/3, 40129 Bologna, Italy}

\emailAdd{georgios.voutsinas@unige.ch}

\abstract{
The Cherenkov Telescope Array Observatory (CTAO) is the next-generation observatory for high energy $\gamma$-ray astronomy with unprecedented sensitivity and accuracy. Accurate estimation and mitigation of systematic uncertainties are crucial for its scientific performance. Atmospheric properties significantly influence both the generation and extinction of Cherenkov light generated by gamma and cosmic rays interacting in the atmosphere. This study provides a detailed analysis of molecular extinction processes, including Rayleigh scattering and molecular absorption, and their impact on the  transmission of Cherenkov light. We examine typical summer and winter behaviour of Rayleigh scattering and seasonal and event-driven variations
of the main absorbing molecules, such as ozone and nitrogen oxides, at the two CTAO array sites. Using simulations, we assess the effects of these variations on image intensity and trigger effective area, particularly during dynamic atmospheric events like stratosphere-to-troposphere transport. Based on our findings, we propose an atmospheric monitoring and calibration strategy to ensure that the CTAO meets its systematic uncertainty requirements, particularly for low-energy gamma-ray observations.
}

\begin{document}
\maketitle

\section{Introduction}

The Cherenkov Telescope Array Observatory (CTAO) is a European research infrastructure for very-high-energy gamma-ray astronomy, currently under construction and later operated by the CTAO ERIC ~\cite{CTAO}. 
The CTAO is made up of two arrays of Imaging Atmospheric Cherenkov Telescopes (IACTs): one in the northern hemisphere, the CTAO-North array, situated at the Roque de Los Muchachos Observatory on La Palma, Canary Islands, Spain, and the CTAO-South array in the southern hemisphere, located near the Paranal Observatory in the Atacama Desert, Chile. The CTAO will host telescopes of three different designs, each of which is optimized for a specific energy range: the  Large-Sized Telescopes (LSTs) of 23~m diameter are optimized for the lowest energies, between 20~GeV and 150~GeV; the Medium-Sized Telescopes (MSTs) of 12~m diameter will cover the core energy range, from 150~GeV up to 5~TeV; while the Small-Sized Telescopes (SSTs) will cover the energy range from 5~TeV to 300~TeV and beyond. The Northern array includes 4 LSTs, currently under construction, and 9 MSTs in the initial Alpha configuration, whereas the Southern array has a larger footprint of about 3 km$^2$ populated by 14 MSTs and 37 SSTs.

 IACTs use the atmosphere as a calorimeter, which needs to be well understood in order to infer the primary gamma-ray energy and direction from the development of the charged component of an electromagnetic (EM) shower the gamma ray induces. IACTs measure gamma rays from cosmic sources in an indirect way. When these gamma rays interact with the atmosphere, they produce an EM shower the development of which is influenced by the atmospheric density profile, which affects both the mean emission height of the Cherenkov light and the distribution of Cherenkov angles. The light that reaches the ground forms a pool of approximately 120~meters in diameter, depending on the altitude of the telescope array, with intensity decreasing exponentially beyond this distance. The image of an EM shower in an IACT camera is approximately elliptical with the major axis pointing toward the direction of the gamma ray source being observed by the array. This shower image provides information not only on the direction to the source, but also on the energy of the primary gamma ray~\cite{Hillas1996}, \cite{hinton2009}. Any changes of the calorimeter properties affect the measurement of the shower energy and its direction as well as the effective area of the instrument.

The transmission of Cherenkov light through the atmosphere is governed by extinction mechanisms, including molecular absorption and scattering processes. Scattering processes include  Rayleigh and particulate scattering. Rayleigh scattering, in this context, is the scattering of light by air molecules. It is often the dominant extinction process and characterized by a strong wavelength dependence, scaling approximately with the inverse fourth power of the wavelength, hence being more effective for shorter wavelengths. 
Particulate scattering (and absorption) occurs due to aerosols and clouds. This process exhibits a weaker and more variable wavelength dependence, influenced by the shape, size and composition of the aerosols. It can become particularly significant during events such as dust intrusions, for example the so-called "calima" events, in which dust from the Saharan desert is transported to the Canary Islands. Molecular absorption is the extinction of light when it traverses gases in the atmosphere that absorb photons. Of particular importance are molecules with an absorption band spectrum in the UV-visible range. In this study we are interested in the wavelength range relevant to the CTAO (280~nm -- 750~nm)~\cite{Mirzoyan:2017}, and more specifically in the range 290~nm -- 550~nm where the efficiency of the telescope's optical transmission is maximum.

This paper investigates the impact of molecular atmospheric absorption %\mg{absorption?}
on the performance of the CTAO. We begin with a brief review of Rayleigh scattering in Sect.~\ref{sec:rayleigh}, followed by an analysis of non-well-mixed atmospheric gases that absorb light in the wavelengths of interest. Specifically, we examine the variations in their mixing ratios at the two CTAO array sites in Sect.~\ref{sec:mixvar}. Subsequently, in Sect.~\ref{sec:molprof}, we propose a methodology for generating Molecular Absorption Profiles (MAPs) using publicly available data. We compare optical depth profiles corresponding to representative atmospheric states. The produced MAPs, adapted to a format compatible with existing simulation tools, will be utilized in simulation studies to assess the effects of their variations on image intensity and trigger effective area in Sect.~\ref{sec:results}. Finally, in Sect.~\ref{sec:disc}, we discuss whether calibrating non-uniformly mixed gases is necessary to ensure that the CTAO systematic uncertainty budget fulfils requirements.

\section{Rayleigh scattering}
\label{sec:rayleigh}

The monochromatic volume molecular scattering coefficient is described by

\begin{equation}
\label{eq:beta}
\beta(\lambda,z) =
\frac{32\pi^3 (n(\lambda,z) - 1)^2}{3\lambda^4 N_s(z)} \cdot
\left( \frac{6 + 3\rho(\lambda,z)}{6 - 7\rho(\lambda,z)} \right) 
\end{equation}

where $N_s(z)$ is the number density of molecules per unit volume at altitude $z$, 
$n(\lambda,z)$ the refractive index of air, and $\rho(\lambda,z)$ the de-polarization factor of air~\cite{Penndorf, Bucholtz}. 
The second factor of Eq.~\ref{eq:beta} is also often called the King factor~\cite{King}, which obtains a slight altitude dependence if water vapour is included. 
Integrating this coefficient from ground level to a given altitude provides the optical depth from the ground up to that altitude. Because $(n-1)$ scales linearly with $N_s$ at a given wavelength, differences in the molecular density profile between atmospheric models~\cite{Munar:2019} will be reflected in the Rayleigh-scattering-induced optical depth. For the purposes of this study, we used the equations from~\cite{Tomasi}, where the water vapour density and CO$_2$ concentration are taken into account in the calculation of the refractive index of the moist air.

The molecular density, water vapour density and refractive index altitude profiles were reconstructed by analysing geopotential, temperature and relative humidity values at various pressure levels, obtained from the European Centre for Medium-Range Weather Forecasts (ECMWF) ERA-5~\cite{ERA5} global climate reanalysis. ERA-5 provides hourly estimates on 37 pressure levels (from 1000~hPa to 1~hPa, roughly corresponding to 45~km of altitude a.s.l.) with a geographical resolution of $0.25^{\circ}\times0.25^{\circ}$. It provides reanalysis data with a latency of 5 days, publicly available via European Union Copernicus service. The CO$_2$ background concentration was retrieved from the Keeling curve~\cite{Keeling,Keeling:2017}. From geopotential, one can get the geopotential height by dividing with the standard Earth's gravitational acceleration. Geopotential heights $\Phi$ are then converted to altitudes above ground $Z$ using the prescription of~\cite{List:1951}: 
\begin{equation}
Z = \frac{R_\phi \Phi}{\left(g_\phi R_\phi/g_0 \right)- \Phi}~,
\end{equation}
where $R_\phi$ and $g_\phi$ are the Earth's radius and local gravity at latitude $\phi$, respectively. An updated version of both can be obtained, for example, from the WGS84 reference ellipsoid~\cite{WGS:1984}.

\begin{figure}
  \centering
   \includegraphics[width=0.99\textwidth]{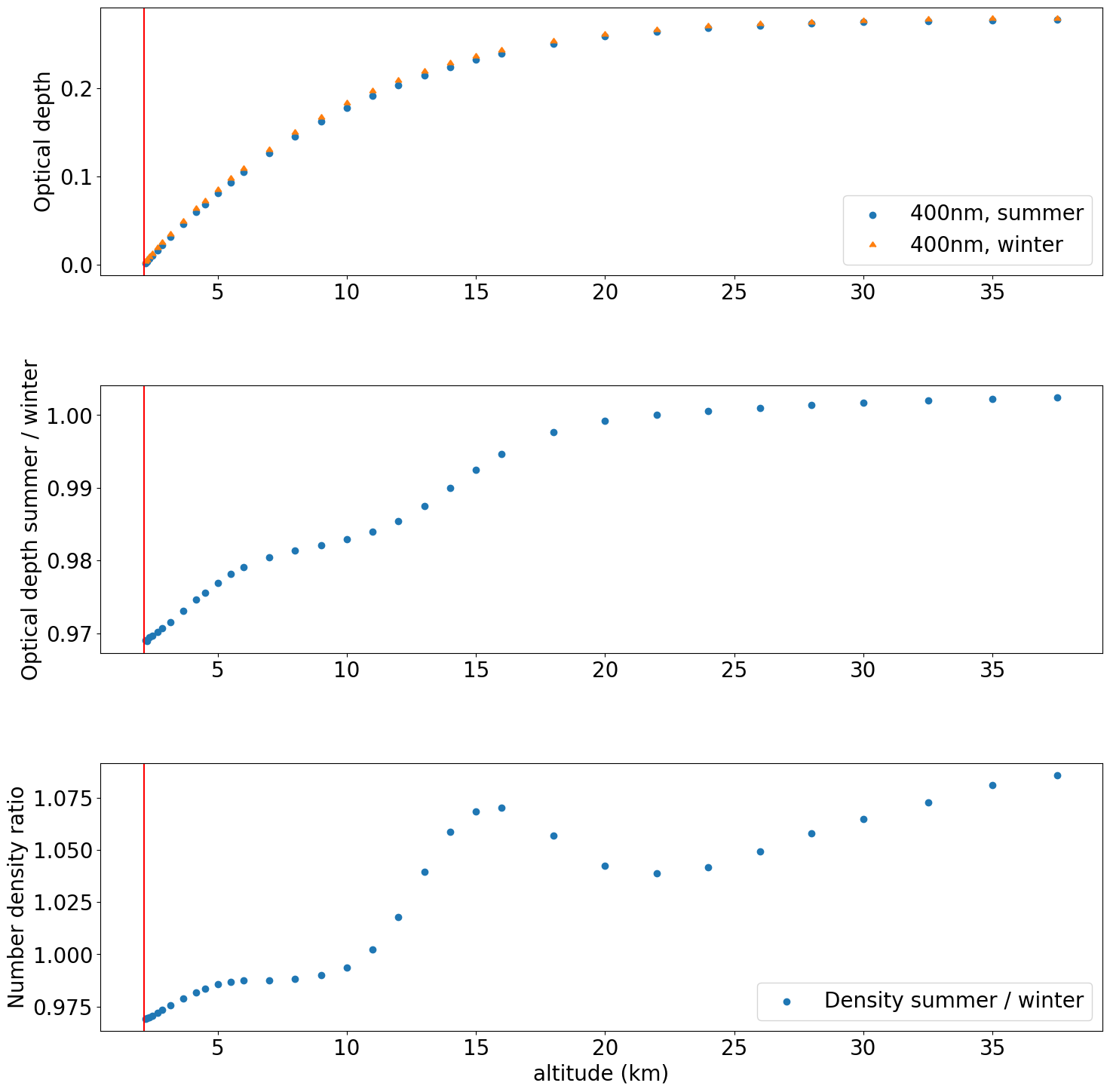}%
   \caption{Comparison of the Rayleigh scattering integrated optical depth at 400~nm as a function of altitude between the two preliminary summer and winter reference profiles of the CTAO-North array site. The red line corresponds to the array site altitude, i.e. 2158~m. The middle panel shows the optical depth ratio between summer and winter. The bottom panel shows the number density ratio. }%
   \label{fig:RS}
\end{figure}

\begin{figure}
  \centering
   \includegraphics[width=0.99\textwidth]{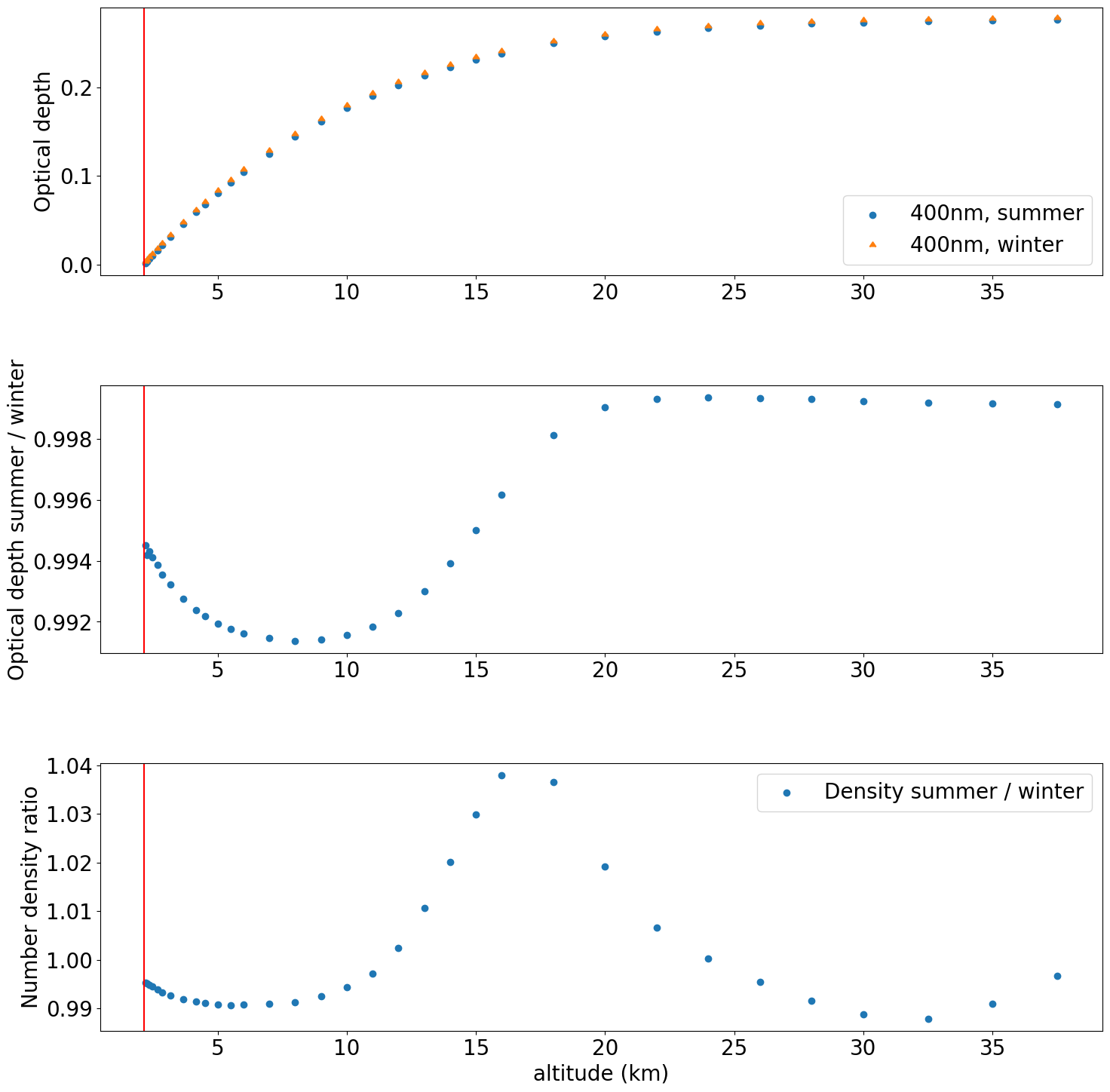}%
   \caption{Comparison of the Rayleigh scattering integrated optical depth at 400~nm as a function of altitude between the two indicative summer and winter reference profiles for CTAO-S. The red line corresponds to the ground level (alt. 2147~m). The middle panel shows the optical depth ratio between summer and winter. The bottom panel shows the number density ratio.}
   \label{fig:RS_paranal}
\end{figure}

Figure~\ref{fig:RS} presents a comparison of altitude profiles for Rayleigh scattering-induced optical depths at 400~nm, derived from indicative summer and winter atmospheric reference profiles for CTAO-North. A similar comparison for CTAO-South is shown in Figure~\ref{fig:RS_paranal}. Figures~\ref{fig:RS} and~\ref{fig:RS_paranal} show the Rayleigh scattering optical depth from the ground to a given altitude. Since the Rayleigh scattering cross-section is proportional to the atmospheric molecular number density, the corresponding optical depth is directly proportional to the integrated number density over the given range. This is a manifestation of the ideal gas law, which dictates that temperature inversely affects molecular number density, thus resulting in a slightly higher tropospheric Rayleigh scattering optical depth during colder seasons. Conversely, during warmer seasons, the tropopause is located at higher altitudes, shifting the scattering profile and slightly increasing the optical depth in the upper troposphere. This effect is reflected in the summer-to-winter optical depth ratio, which approaches unity around and above the tropopause altitude.

\section{Molecular absorption}

In the wavelength range relevant to this study, ozone is the primary light-absorbing molecule. It exhibits three characteristic absorption bands: the Hartley band (200--310~nm), which is the strongest; the Huggins band (310--350~nm), which is weaker; and the Chappuis band (400--650~nm), with two absorption maxima at 575~nm and 603~nm \cite{Bogumil2003}. Due to the telescope's design requirement for insensitivity to wavelengths below 290~nm, only the portion of the Hartley band above this threshold is relevant for this analysis.
Nitrogen oxides (NO$_x$) also absorb in the UV and visible ranges, but their effect is expected to be smaller than that of ozone.
In contrast, molecular oxygen and water vapour are not considered in this study. Molecular oxygen absorbs strongly below 250~nm---outside the telescope's sensitivity range---and exhibits weaker features at 630~nm and 690~nm (the Fraunhofer $\gamma$ and B bands), where the telescope’s efficiency is very low. Water vapour features a dominant absorption band at 720~nm, which is likewise a region of negligible telescope sensitivity. Consequently, absorption by both molecular oxygen and water vapour is deemed insignificant for the purposes of this analysis.

Ozone and NOx are non-well-mixed gases in the atmosphere, which means that they are non-uniformly distributed.  
Tropospheric ozone is predominantly produced through photochemical oxidation of CO and hydrocarbons catalysed by HOx and NOx~\cite{Monks:2015}, followed by Stratosphere-to-Troposphere Transport (STT)~\cite{Regener:1938,Olsen:2004}, generated by tropopause folds, gravity wave breaking and deep convection, and where falling air masses can bring ozone-rich air parcels from the stratosphere into the troposphere. Also biomass burning has been recently identified as a major contributor to tropospheric ozone~\cite{Bourgeois:2021}. Whereas chemical production  contributes about eight times more to global tropospheric ozone than STT processes~\cite{Monks:2015}, the latter contribute to more than three fourths of the interannual variability~\cite{Voulgarakis:2010}. In particular, at CTAO-S, background ozone is mainly driven by stratospheric intrusions rather than photochemical production~\cite{Anet:2017}. 
Ozone is highly reactive chemically, which results in a short lifetime. At remote locations, ozone is mainly destroyed by photolysis~\cite{Ayers:1992}. The lifetime of ozone in the free troposphere ranges from about a week or so in lower altitudes to a few months in the upper troposphere~\cite{Archibald:2020}. 
Ozone and ozone precursors are regularly exported from their emission source to
receptor regions far downwind on a regional, inter-continental and even hemispheric scale. Pollution plumes not
only travel from North America to Europe, or from East Asia to North America, but can also circle the globe~\cite{Monks:2015}.  
At the Canary Islands, increased levels of tropospheric ozone are associated with air masses travelling above 4~km altitude from North America over the North Atlantic Ocean, and, during summer, with STT processes from regions neighbouring the Canary Islands~\cite{Cuevas:2013}.
An increase of tropospheric ozone in the range from (1.5--3)\% per decade has been registered over the Canary Islands, influenced by the evolution of the North Atlantic Oscillation (NAO)~\cite{Cuevas:2013,Gaudel:EA2018a,VanMalderen:2025}. 

For Northern Chile, \cite{Le:2024} find that the influence of the El Ni\~no-Southern Oscillation (ENSO), and particularly changes of ENSO cycles due to global warming, on tropospheric ozone above continents is insignificant, contrary to the case of tropospheric ozone above the Pacific ocean. 

In the stratosphere, ozone is produced mostly through photochemical reactions involving UV sunlight and molecular oxygen. Ozone production depends on the solar flux, which varies with season and geographical latitude, while its destruction is influenced by solar UV radiation and anthropogenic pollutants such as chlorofluorocarbons (CFCs) or NOx. Dynamic transport phenomena, which can move large quantities of these gases vertically or horizontally within the atmosphere, can rapidly change significantly the ozone mixing ratios~\cite{copernicus_stt}. Of particular interest for this study are the STT events.
All the above reasons lead to large spatial and vertical gradients. 

On remote locations, NOx mixing ratios are affected by anthropogenic factors, such as aircraft emissions~\cite{Wang:2022}, and natural processes, such as lightning~\cite{schumann:2007} as well as peroxyacetyl nitrate~\cite{Archibald:2020}, a temporary reservoir species for NOx that is thermally unstable and which is formed primarily in the urban atmosphere from where it can be transported over long distances in the free troposphere, facilitating ozone production in the remote atmosphere. In the context of this study, NOx play a dual role: they not only affect ozone mixing ratios but also absorb light directly. This motivated our focus on the variability of non-uniformly mixed gases and their potential impact on the scientific performance of the CTAO. Accordingly, we specifically target ozone and NOx in this work.

\subsection{Data sources}

For the purposes of this study, we use data from the major global Data Assimilation Systems (DAS). The ozone mass mixing ratio can be retrieved either from the ECMWF ERA-5 dataset or from the ECMWF Atmospheric Composition Reanalysis 4 (EAC4) dataset~\cite{EAC4}. EAC4 provides a full representation of ozone chemistry, while ERA-5 has a more simplified one (which, however, includes the ozone-hole-related chemistry). On the other hand,  ERA-5 features a finer grid with a geographical resolution of $0.25^{\circ}\times0.25^{\circ}$, instead of $0.75^{\circ}\times0.75^{\circ}$ offered by EAC4. An additional reason to choose ERA-5 is that its near real time reanalysis data are considered a data source for the CTAO atmospheric characterization pipeline, making the application of ozone calibration, if finally required, more straightforward. The mass mixing ratios of NOx are retrieved from EAC4 which includes detailed catalogues of atmospheric composition based on more advanced chemical modelling compared to ERA-5.
The GDAS analysis dataset ds083.2~\cite{ds083.2} that provides analysis data with a latency of about 2 hours, is used for cross validation.  

\subsection{Mixing ratios variations}
\label{sec:mixvar}

This study examines the seasonal and event-driven variations in ozone and nitrogen oxides at various pressure levels over the CTAO sites. For context, the nighttime surface atmospheric pressure ranges from 765 to 800 hPa at CTAO-North~\citep{GaugWS:2024}, and from 778 to 790 hPa at CTAO-South, which can help interpreting the pressure-level plots more effectively.

\subsubsection{Ozone}

\begin{figure}
    \centering
    \includegraphics[scale=0.6]{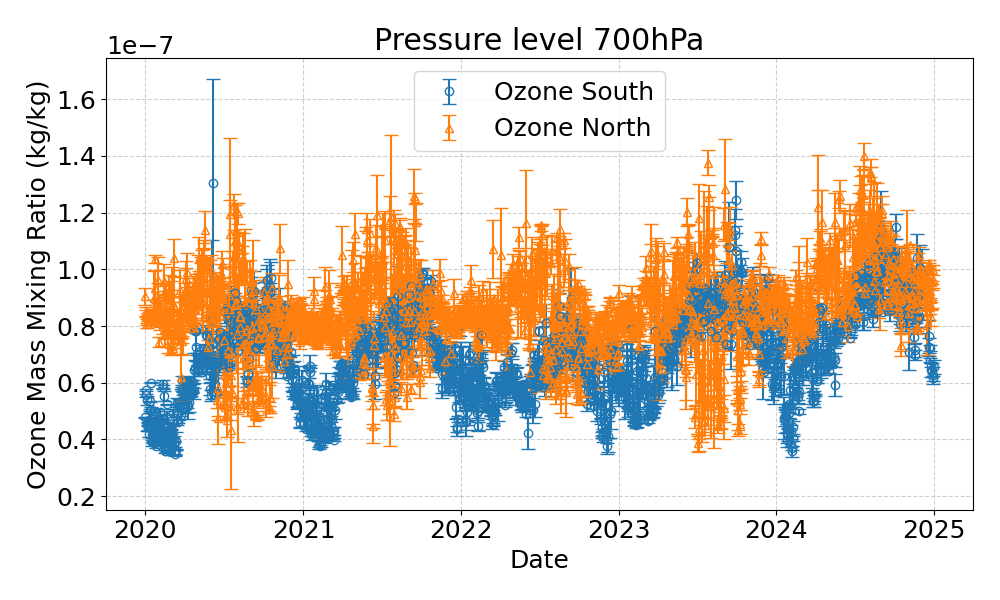}
    \caption{Nighttime ozone mass mixing ratio over the Southern and Northern CTAO sites, in the time period 2020-2024, at a pressure level of 700~hPa.}
    \label{fig:ozone_seasonal}
\end{figure}

The ozone mixing ratio exhibits a regular seasonal cycle. Additionally, it can be affected by horizontal or vertical transport of ozone masses, in particular STT events~\cite{Olsen:2004}.  Figure~\ref{fig:ozone_seasonal} shows the changes in nightly ozone mass mixing ratio over five years (2020-2024) at a pressure level of 700~hPa (corresponding to roughly 1~km above ground) over the two CTAO sites. The seasonal variation is clearly visible. Each point in Figure~\ref{fig:ozone_seasonal} corresponds to the night time average for a given date. The error bars depict the standard deviation, calculated from the hour-to-hour variation within a single night. Points where the mixing ratio is significantly higher than that of the neighbouring data points can possibly be attributed  to dynamic transport events~\cite{Anet:2017}.

Dynamic transport events can be either deep—reaching the Planetary Boundary Layer (PBL)—or shallow, affecting only the upper troposphere. While we do not attempt to classify the events in this study (as that would require estimating the PBL thickness, which is beyond the scope of this work), we aim to identify such events at various pressure levels.

\begin{figure}
    \centering
    \includegraphics[scale=0.6]{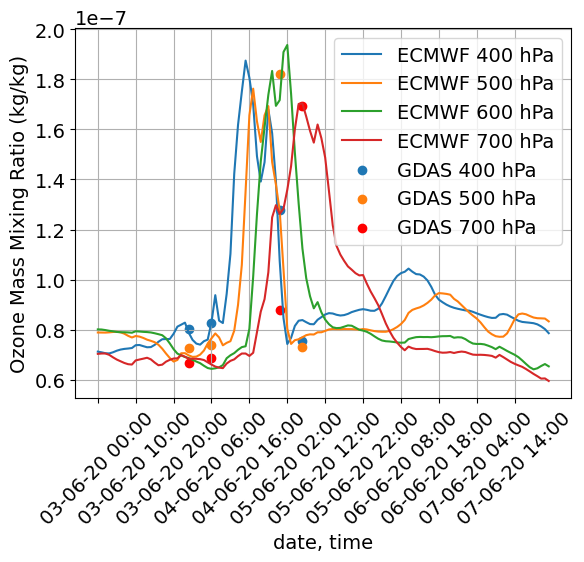}
    \caption{Time - altitude evolution of a STT ozone mass transport event that took place at CTAO-South array site in June 2020.}
    \label{fig:stt}
\end{figure}

At the 700 hPa pressure level, as shown in Figure~\ref{fig:ozone_seasonal}, only one clear event is identifiable: it occurred in June 2020 over the CTAO-South site. A more detailed view of the event’s evolution is presented in Figure~\ref{fig:stt}. During this event, the ozone mixing ratio increases rapidly—by more than a factor of two. The entire event spans approximately two days, with ozone-enriched air persisting at each pressure level for several hours before continuing its descent. This event is detected in both the ECMWF and GDAS datasets. The two data assimilation systems show good agreement in their predictions, although ECMWF provides significantly higher temporal resolution.

\begin{figure}
    \centering
    \includegraphics[width=0.99\textwidth]{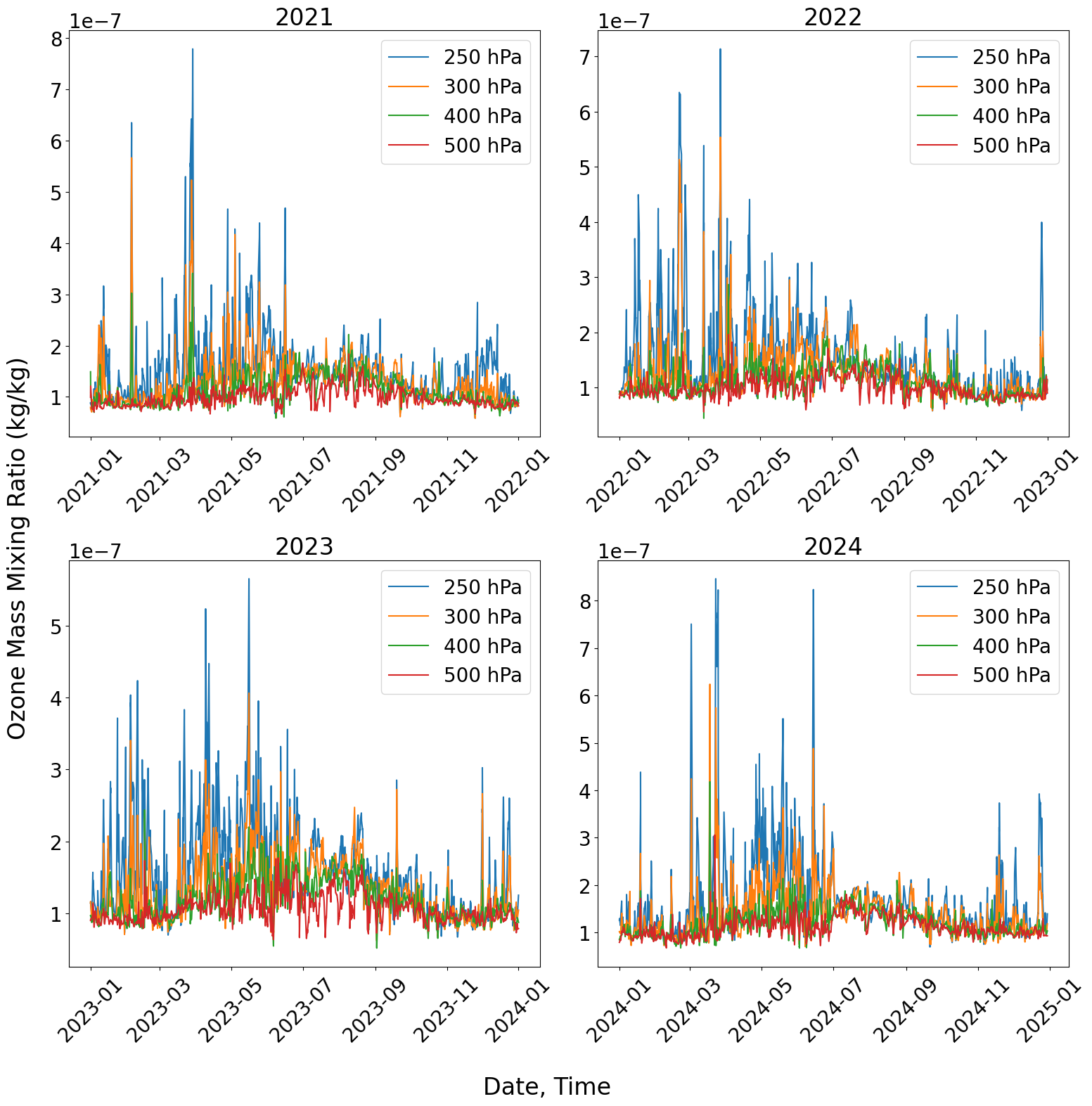}
    \caption{Ozone mass mixing ratio over the CTAO-North site during the years 2021-2024.}
    \label{fig:stt_shallow_north}
\end{figure}

At lower pressure levels, the frequency of dynamic transport events is higher. Figure~\ref{fig:stt_shallow_north} presents the ozone mixing ratios over the CTAO-North site for four consecutive years. During winter and spring, these events occur frequently—every few days—and are often striking in magnitude, with ozone mixing ratios increasing by up to a factor of six. The observed variation in the frequency of STT events throughout the year agrees with the results presented in~\cite{copernicus_stt}. One can observe that the ozone mixing ratio at 500~hPa is relatively smooth. The spikes occur mostly at 250-300~hPa.

\begin{figure}[ht]
    \centering
    \begin{subfigure}{0.45\textwidth}
        \centering
        \includegraphics[width=\linewidth]{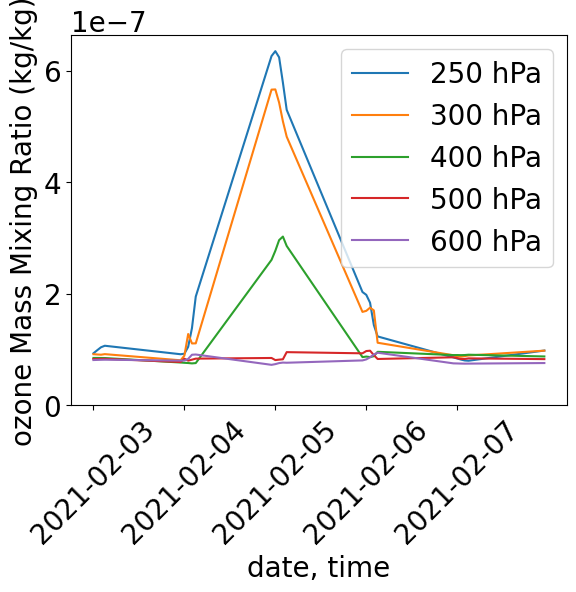}
        \caption{}
        \label{fig:subfig1}
    \end{subfigure}
    \hfill
    \begin{subfigure}{0.45\textwidth}
        \centering
        \includegraphics[width=\linewidth]{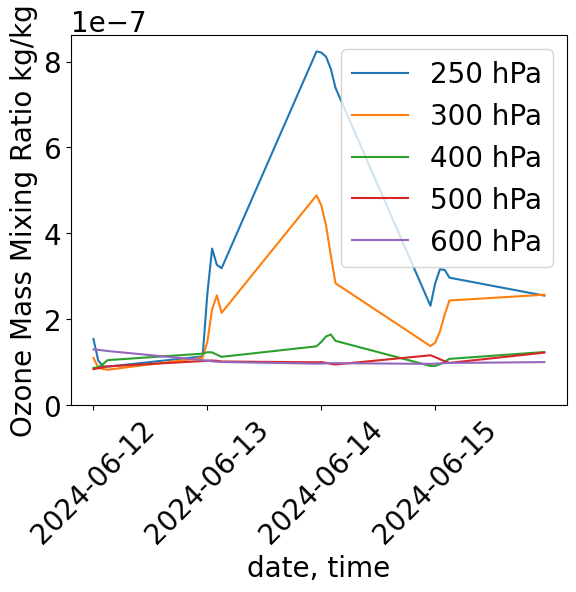}
        \caption{}
        \label{fig:subfig2}
    \end{subfigure}
    \caption{Time–altitude evolution of two stratosphere-to-troposphere transport (STT) events observed above the CTAO-North site. (a) A strong shallow ozone mass transport event in February 2021, identified as one of the most intense STT events during 2020–2024, but coinciding with a snowstorm, which prevented telescope operation. (b) An STT event in June 2024 under favorable weather conditions, during which the telescopes would be operational.}
    \label{fig:stt_north_feb_2021}
\end{figure}

Figure~\ref{fig:stt_north_feb_2021} shows a shallow STT event that occurred in February 2021 over CTAO-North. As in the previous case, the event unfolds over the course of several days. However, unlike the event depicted in Figure~\ref{fig:stt}, pressure levels at or above 500 hPa are not affected. In contrast, the impact on ozone mixing ratios at lower pressure levels is significantly more pronounced, with concentrations increasing by a factor of six. These observations highlight the need to examine the potential impact of both deep and shallow dynamic transport events on CTAO's performance.

\begin{figure}
    \centering
    \includegraphics[width=0.99\textwidth]{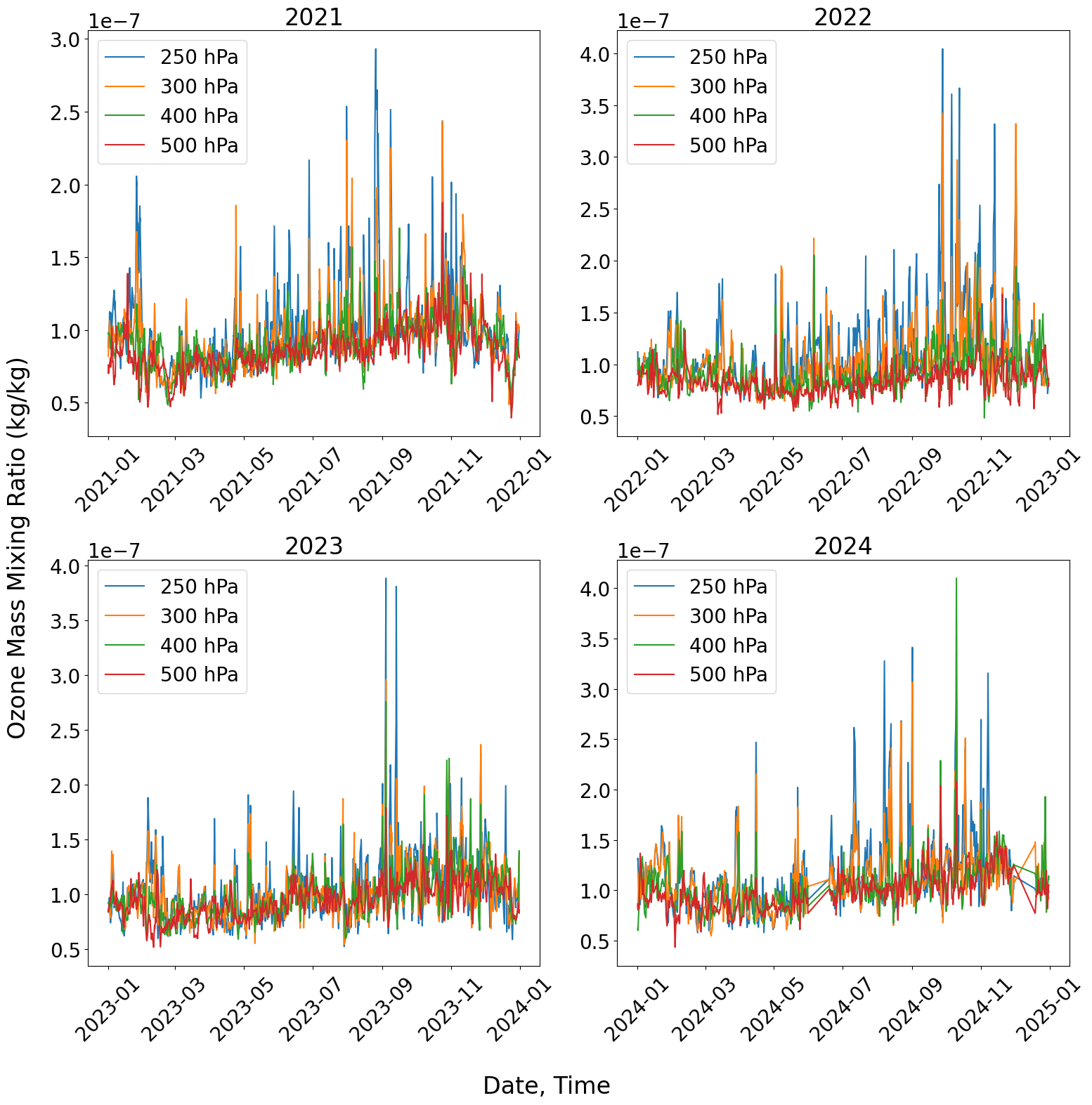}
    \caption{Ozone mass mixing ratio over the CTAO-South during the years 2021-2024.}
    \label{fig:stt_shallow_south}
\end{figure}

Similar trends are observed at CTAO-South (Figure~\ref{fig:stt_shallow_south}). Shallow transport events occur more frequently during the austral spring which is consistent with~\cite{Anet:2017} who find the strongest STT events in October at Cerro Tololo. 
However, their absolute magnitude is smaller compared to those observed at CTAO-North.

\subsubsection{Nitrogen oxides}

\begin{figure}
  \centering
   \includegraphics[scale=0.55]{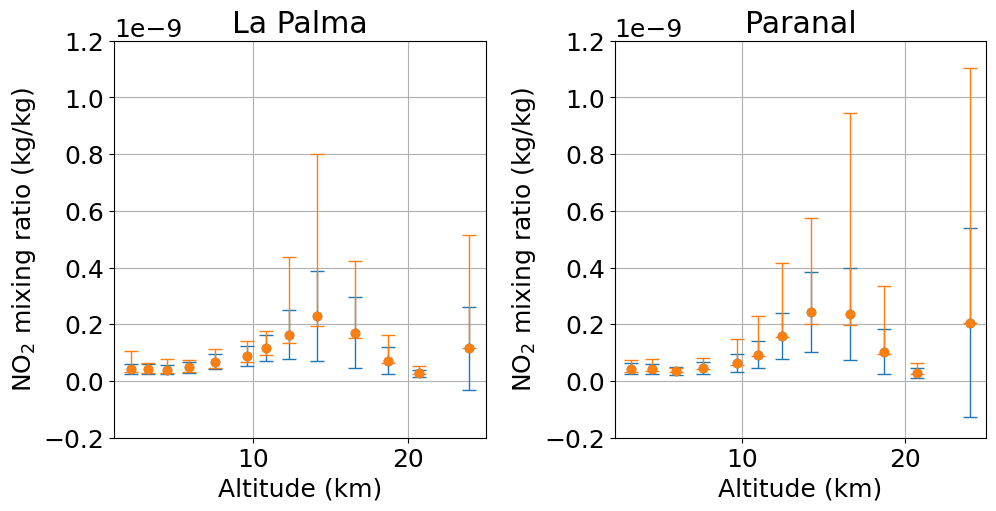}%
   \caption{Nitrogen dioxide mixing ratio at different altitudes over the two CTAO sites (left: CTAO-North, right: CTAO-South). Error bars represent the interquartile range.} 
   \label{fig:no2}
\end{figure}

Nitrogen oxides also absorb light in the wavelength range of interest. Nitrogen dioxide (NO$_2$) is expected to contribute to Cherenkov light extinction, albeit to a lesser extent than ozone. We analyzed EAC4 datasets for nighttime conditions in 2023 for both CTAO sites. The altitude profile of the NO$_2$ mixing ratio is shown in Figure~\ref{fig:no2}, with CTAO-North on the left and CTAO-South on the right. The blue error bars represent the interquartile range observed throughout the year at each altitude. The altitude range considered is from 2.5~km to 25~km a.s.l.. Local maxima are observed around the tropopause, originating from sources such as aircraft emissions~\cite{Wang:2022} and pollutants transported upward via convection.

\begin{figure}
  \centering
   \includegraphics[scale=0.55]{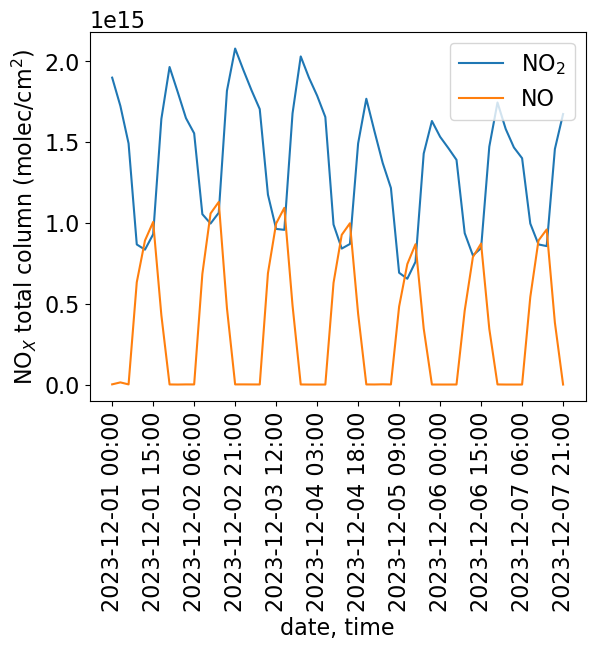}%
   \caption{Total column of NOx during the first week of December 2023 at CTAO-North array site.}
   \label{fig:no}
\end{figure}

Nitrogen monoxide (NO) mixing ratios are significantly reduced during nighttime because most NO is oxidized to NO$_2$, and the lack of photolysis prevents the regeneration of NO, shifting the equilibrium toward NO$_2$~\cite{Finlayson-Pitts:2000}. 
Figure~\ref{fig:no} illustrates the diurnal variations of the total column of NO$_X$ during one representative week—specifically, the first week of December 2023—at the CTAO-North array site. This week was chosen for illustrative purposes only and is not unique; similar patterns are expected during other weeks and at both the Northern and Southern CTAO sites. It should be noted that discrepancies between model predictions and direct observations of the NO$_2$ total column can be as large as a factor of two~\cite{Shah:2023}. Consequently, we adopt a factor of two as a conservative estimate of the uncertainty in the reported total column values. Additionally our analysis is in contradiction with ~\cite{Garcia:2021}, in which the total column of NO observed at Tenerife oscillates around $4\times 10^{15}$~cm$^{-2}$ and NO$_2$ around $3\times 10^{15}$~cm$^{-2}$. As expected, the NO total column approaches zero during nighttime. Therefore, we assume that the effect of nitrogen monoxide on the performance of CTAO can be safely neglected.

\subsection{Production of molecular absorption profiles}
\label{sec:molprof}

To estimate the effect of molecular absorption and its variations on the CTAO science performance, we generated the corresponding MAPs and converted them into a format compatible with the CTAO simulation software~\cite{ctadirac}. The information contained in a MAP is the integrated optical depth (or transmission) per altitude bin and per wavelength. To produce a MAP, we started with the mass mixing ratio of the target molecule, retrievable from ECMWF datasets in units of kg/kg. 
The conversion of the mixing ratio to the number density, for example for ozone, was performed using Eq.~\ref{number_density}:
\begin{equation}
    \label{number_density}
    n_{O_3}(h) = \frac{\chi_{O_3}(h) \times \rho_{\mathrm{atm}}(h) \times N_{A}} {M_{O_3}},
\end{equation}

where $n_{O_3}(h)$ is the ozone number density as a function of altitude $h$, $\chi_{O_3}(h)$ the ozone mass mixing ratio, $\rho_{\mathrm{atm}}(h)$ the atmospheric density as a function of altitude $M_{O_3}$ the ozone molar mass and $N_A$ the Avogadro number. The atmospheric density profile can be calculated using the same ERA-5 dataset that we use to retrieve the ozone mass mixing ratio, or alternatively, it can be obtained from a reference atmospheric model.

Multiplying $n_{O_3}(h)$ by the absorption cross section $\sigma(\lambda)$ yields the extinction coefficient $\alpha(h, \lambda)$, expressed as a function of altitude $h$ and wavelength $\lambda$. The absorption cross sections for the various molecules were obtained from the HITRAN database~\cite{hitran} for nitrogen dioxide, while for ozone from a combination of HITRAN (243~nm to 346~nm at 293 $^\circ$K) and Ref.~\cite{Gorshelev} for the rest of the wavelength range. The optical depth from the array site level to a given altitude was calculated by integrating $\alpha(h, \lambda)$ up to that altitude. Consequently, by performing this integration across a specified set of altitudes, we produced the MAP. The software used to generate MAPs is described in Ref.~\cite{mdps}. 

As mentioned in the introduction, three main processes influence the propagation of Cherenkov light through the atmosphere. Therefore, extinction profiles that account for all these processes must be produced.  The addition of the profiles is performed in accordance with the Beer-Lambert law, Eq.~\ref{beer}:

\begin{equation}
    \label{beer}
    \tau = \sum_{1}^{n} \tau_i
\end{equation}
where $\tau$ is the overall optical depth and $\tau_i$ the optical depth contribution of each process.

For the purposes of this study, we produced Molecular Extinction Profiles (MEPs), defined as the combination of MAPs with Rayleigh scattering extinction profiles. The aerosol component is not included (see Sect.~\ref{sec:shortcuts}).
For CTAO-South, we discuss two seasonal profiles derived from five years (2019–2023) of night-time data for the months of January and February (austral summer) and July and August (austral winter), as well as one extreme profile corresponding to the period from June 4 to June 6, 2020, during the STT event shown in Figure~\ref{fig:stt}.
To isolate the impact of absorbing molecule variability from Rayleigh scattering effects, all MEPs in this study use the same average Rayleigh scattering extinction profile.
For CTAO-North, we compare the average winter MEP with the profile that includes the STT event that occurred on February 5, 2021.

\subsubsection{Optical depths for different profiles}
\label{od_results}

\begin{figure}
  \centering
   \includegraphics[width=0.9\textwidth]{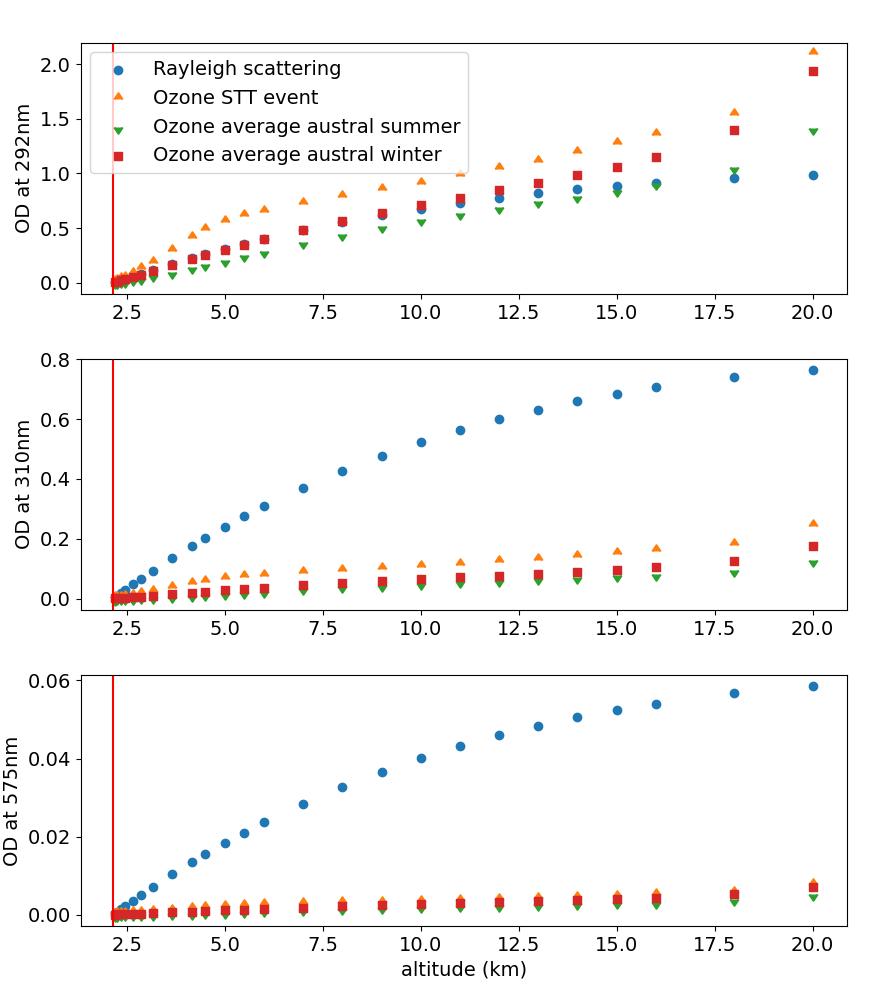}%
   \caption{Comparison of the cumulative optical depth as a function of altitude for Rayleigh scattering and ozone absorption for three different wavelengths: top for 292~nm, within the Hartley bands where the ozone absorption is most pronounced; middle for 330~nm, within the Huggins bands; and bottom for 575~nm, which corresponds to a local maximum of the Chappuis bands. The vertical red line at an altitude of 2160~m corresponds to the ground level at the CTAO-South site.}%
   \label{fig:od_comp}
\end{figure}

Figure~\ref{fig:od_comp} illustrates the relative contributions of the two primary molecular extinction processes for the CTAO-South site. It compares an average Rayleigh scattering optical depth profile, representative of conditions over this site, against the following two ozone profiles: one corresponding to the STT event, one to austral winter, and one to austral summer. At shorter wavelengths, within the Hartley bands, the ozone absorption cross section reaches its maximum values, leading to ozone-related extinction that is comparable to or even exceeds (during austral winter) Rayleigh scattering. Conversely, at longer wavelengths, Rayleigh scattering becomes the dominant extinction process. Seasonal variations in ozone-related optical depths are evident, with stronger light extinction observed during the austral winter in the altitude range between 5~km and 12~km a.s.l., where an extensive air shower (EAS) is typically developed. Furthermore, during an STT event, the ozone-related contribution to the optical depth in the troposphere increases significantly, exceeding its average value by a factor of two or more.

\section{Impact on CTAO performance}
\label{sec:results}

\begin{figure}
  \centering
   \includegraphics[width=0.99\textwidth]{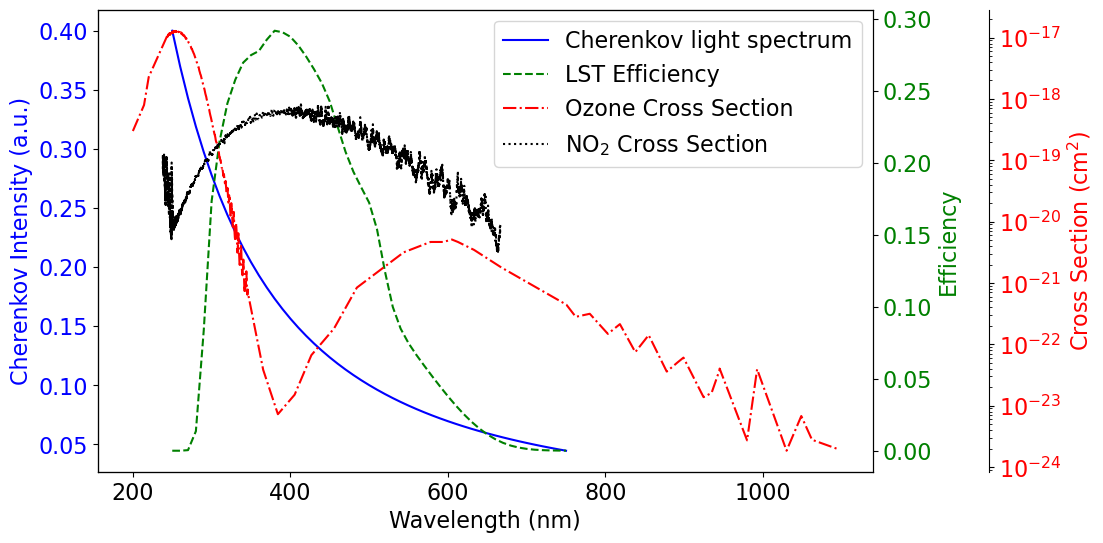}%
   \caption{Cherenkov light spectrum (blue), LST telescope efficiency, and atmospheric absorption cross sections (red: ozone, black dotted: NO$_2$) as a function of wavelength.}
   \label{fig:testeff}
\end{figure}

Variations in the mixing ratios of atmospheric absorbing molecules modify the atmospheric optical depth, thereby affecting the amount of Cherenkov light that reaches the ground. This, in turn, impacts both the energy reconstruction and the effective area of the CTAO. A first-order estimate of these effects on the  Cherenkov light registered by a telescope, without resorting to full air-shower simulations, can be obtained by convolving the atmospheric absorption profile with the Cherenkov emission spectrum and the wavelength-dependent optical throughput of the telescope.
This approach is implemented in the \textit{testeff} program, which is distributed as part of the \textit{sim\_telarray} software package~\cite{simtel}. The \textit{testeff} tool propagates Cherenkov light through the atmosphere and the telescope over a user-defined wavelength range, starting from a specified emission altitude. It computes the Cherenkov light detection efficiency by multiplying wavelength-dependent efficiency tables of the individual telescope components with the corresponding atmospheric optical depth, and subsequently convolves the result with the Cherenkov light spectrum.
Figure~\ref{fig:testeff} illustrates this procedure and highlights the interplay between the Cherenkov emission spectrum, telescope optical efficiencies, and atmospheric absorption processes. Although Cherenkov radiation is intrinsically strongest at short wavelengths, the effective detected signal is significantly reduced by strong ozone absorption in the UV and is further modulated by the instrumental efficiency curves of the telescopes. Additional absorption features introduced by nitrogen dioxide in the visible band further shape the detected spectrum.
An increase in atmospheric extinction is therefore expected to predominantly affect the lowest energies, reducing image intensity and the effective area near the energy threshold of each telescope. Since low-energy showers traverse a larger part of the atmosphere, we provide testeff with the parameters of LST, as LSTs are the CTAO telescope class optimised for low-energy gamma-ray observations.

For the CTAO-South, we examined the two cases illustrated in Figure~\ref{fig:od_comp}, i.e., the STT ozone profile versus the austral winter one. We considered them both independently (as MAPs) and in combination with an average Rayleigh scattering extinction profile (as MEPs). This study assumed Cherenkov light arriving from zenith, in a wavelength range of 280~nm to 750~nm and a starting altitude corresponding to an atmospheric thickness equal to 200~g/cm$^2$. For CTAO-North, we compared the average winter ozone profile with that observed during the stratosphere-to-troposphere transport (STT) event on February 5, 2021. The results are presented in Table~\ref{tab:testeff}. 

\begin{table}
    \centering
    \begin{tabular}{|c|c|c|}
        \hline
        extinction profile & telescope and atmosphere efficiency & efficiency drop (\%) \\
        \hline
        ozone STT South & 
        0.1654
        %0.13916 
        & \multirow{2}{*}{1.3}\\
        \cline{1-2}
        ozone austral summer & 
        0.1676
        %0.140659 
        & \\
        \hline
        MEP STT South & 
        0.1255
        %0.105323 
        & \multirow{2}{*}{0.7} \\
        \cline{1-2}
        MEP (austral summer) & 
        0.1264
        %0.106056 
        & \\
        \hline
        MEP STT North & 
        0.1225
        %0.102827 
        & \multirow{2}{*}{2.1} \\
        \cline{1-2}
        MEP North winter & 
        0.1251
        %0.104954 
        & \\
        \hline
    \end{tabular}
    \caption{LST efficiency to detect Cherenkov light, integrated in the wavelength range 280--750 nm, for different extinction profiles. The optical transmission of the various telescope components has been taken into account.}
    \label{tab:testeff}
\end{table}

The results show that during the STT event at CTAO-South, the Cherenkov light detected by the LSTs decreased by approximately 0.7\%. In contrast, a larger decrease of about 2\% was bserved at CTAO-North. Finally, we estimated the relative contribution of ozone-related extinction compared with the dominant Rayleigh-scattering-induced contribution and find it to be approximately 15\%. Tests carried out at higher zenith angles (not shown in the table) indicate that the magnitude of the effect is of a similar size or even slight smaller at higher zenith angles.

A similar analysis was conducted for NO$_2$. We evaluated the programme output with and without the inclusion of a NO$_2$ MAP. The overall impact on Cherenkov light transmission was found to be $\leq$ 0.02\%.

\subsection{Full simulations}

The decrease in registered Cherenkov light observed in the MEPs that include STT events is expected to introduce a bias in energy reconstruction and the effective area if no ozone correction is applied. To estimate the magnitude of these effects, we examine the impact of different MEPs on image intensity and trigger effective area using dedicated full simulations. Variations in image intensity and trigger effective area are then used as first-order proxies for the corresponding biases in reconstructed energy and effective area, respectively. The CTAO-North array was simulated using the Alpha configuration layout, consisting of four LSTs and nine MSTs. The CTAO-South array was simulated with a modified Alpha~\cite{alpha} configuration layout, which includes four LSTs, 14 MSTs, and 42 SSTs. A total of 300 million gamma-ray showers were generated with CORSIKA~\cite{Corsika} using the PROD6 configuration~\cite{prod6} at a zenith angle of 20 degrees. Each CORSIKA dataset is processed with \textit{sim\_telarray}, applying the corresponding MEP profile in each case. \textit{Sim\_telarray} models the response of the CTAO telescopes, including the optics, camera electronics, and trigger systems, under realistic observing conditions.

\begin{figure}
  \centering
   \includegraphics[width=0.99\textwidth,trim={0cm 0cm 0cm 0cm},clip]{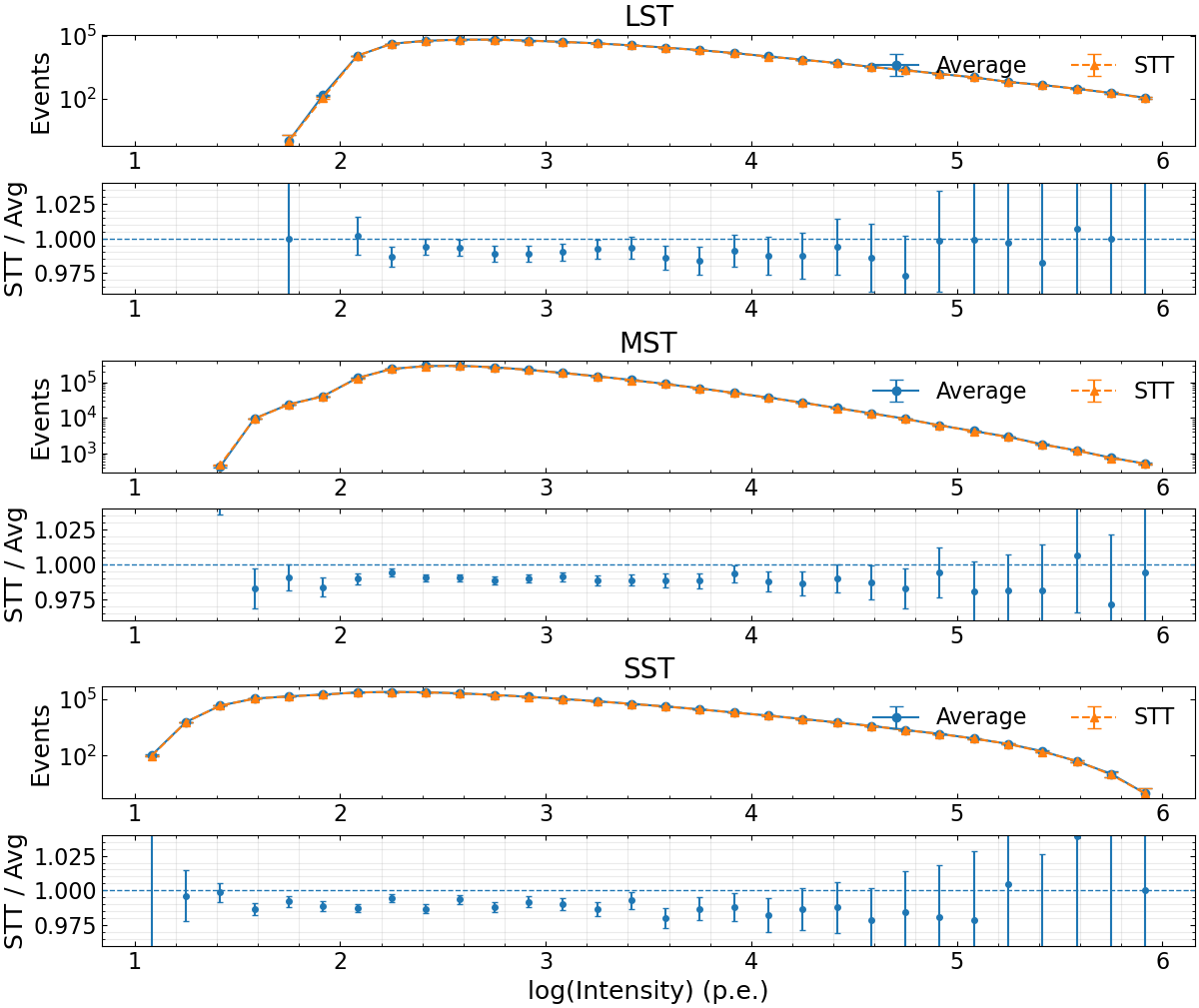}%
   \caption{Distributions of the logarithm of the image intensity for triggered stereo events from CTAO-South array simulations, shown separately for LSTs, MSTs, and SSTs. For each telescope type, the main panels compare simulations produced with the average molecular extinction profile and with the STT molecular profile, while the lower panels show the bin-by-bin ratio between the two distributions. Error bars represent statistical uncertainties.}
   \label{fig:II_south}
\end{figure}%

\begin{figure}
  \centering
   \includegraphics[width=0.99\textwidth,trim={2cm 0.5cm 2cm 1cm},clip]{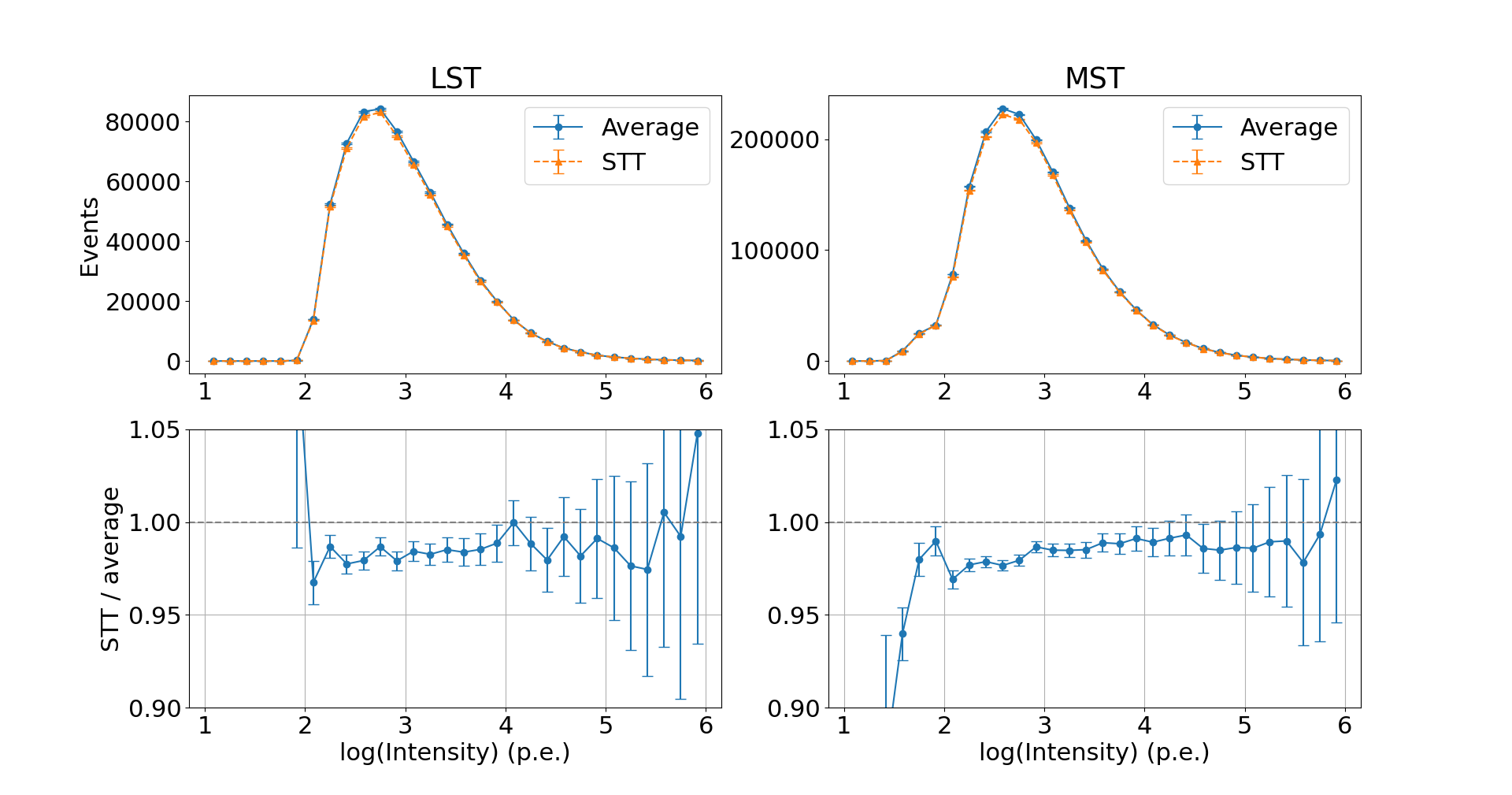}%
   \caption{Distribution of the logarithm of image intensities for triggered stereo events from simulations of the CTAO-North array under two different molecular extinction profiles.  The error bars represent statistical uncertainties. The left panel shows the results for LST, while the right for MST. The bottom panel shows the ratio between the two profiles.}%
   \label{fig:II_north}
\end{figure}%

Figure~\ref{fig:II_south} presents a comparison of simulated stereo-triggered events image intensities, per telescope type, between an MEP produced with an average ozone extinction profile and one featuring the STT event for CTAO-South. The stated intensity is the sum of the intensities of all triggered telescopes of  the same type in a given event. The lower panels show the ratio of the simulated image intensities. Within the range of 100 to 100,000 photoelectrons, this ratio consistently remains below one, indicating that images recorded during STT events are slightly dimmer by approximately 1~\%. Outside this range, increased statistical uncertainties hinder definitive conclusions. A similar comparison for the Northern array is shown in Figure~\ref{fig:II_north}. It is important to recall the differences between the two STT events. The Southern STT event featured a doubling of the ozone mixing ratio that extended to higher pressure levels ($\geq$ 700~hPa), whereas the Northern STT showed a sixfold increase in ozone mixing ratio, which was already halved at 400~hPa and is completely absent at 500~hPa. As a result, the Northern STT is expected to affect more the low energy showers. This is evident in Figure~\ref{fig:II_north}, where the lower intensity images are affected more than the brighter ones. For images with less than 1000 photoelectrons, the effect appears to be larger than 2~\%.

\begin{figure}
  \centering
   \includegraphics[width=0.99\textwidth,trim={3cm 0.5cm 4.5cm 3cm},clip]{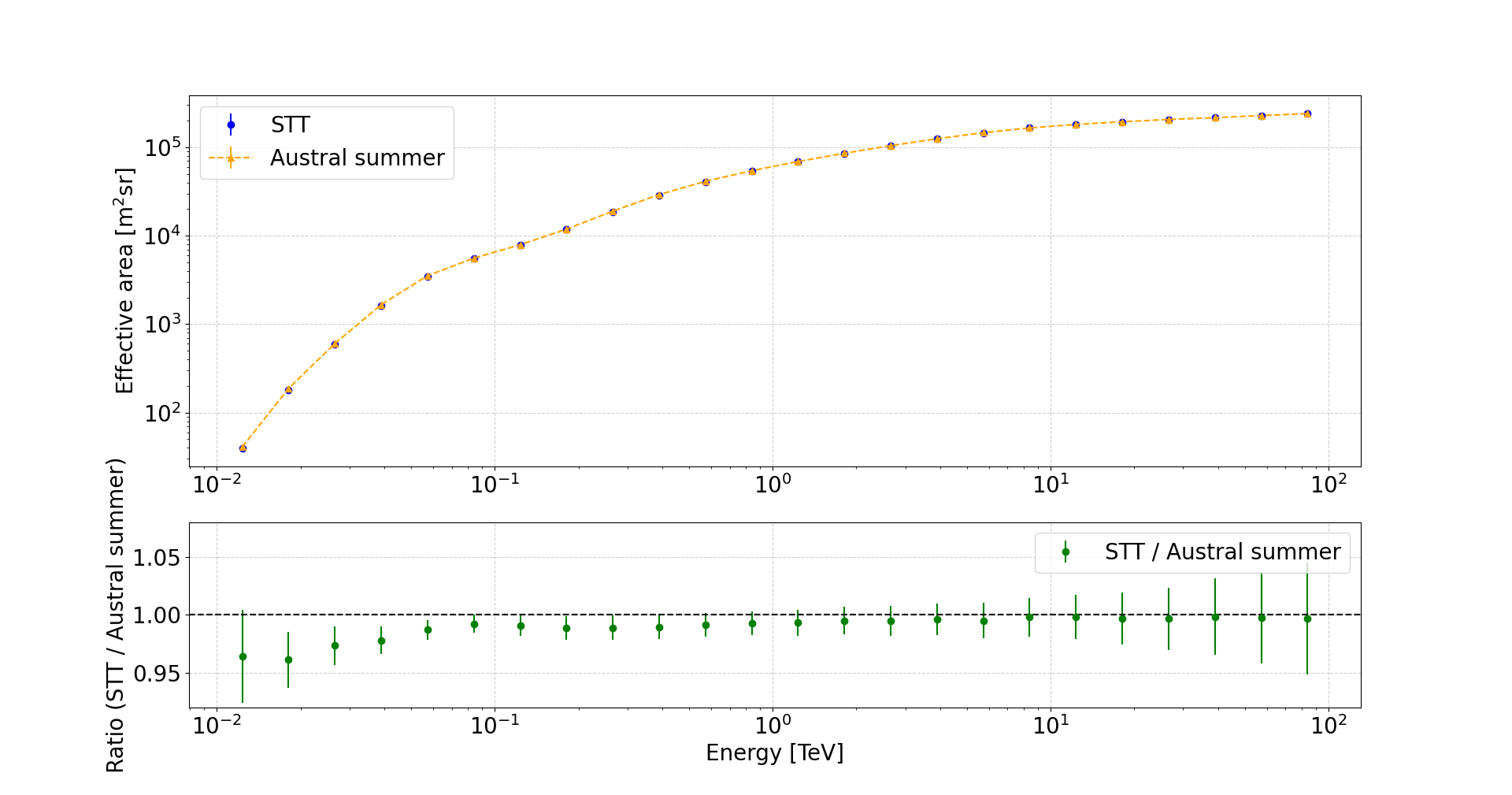}%
   \caption{Top panel: stereo trigger effective areas for the CTAO-South array under two different molecular extinction profiles: after incorporating an STT event or the average austral summer ozone profile. Bottom panel: ratio of the two trigger effective areas. Note that the uncertainties at energies above 1 TeV are affected by the intrinsic uncertainties of the effective areas themselves, as a consequence of limited simulation statistics.}%
   \label{fig:ea}
\end{figure}%

\begin{figure}
  \centering
   \includegraphics[width=0.99\textwidth,trim={3cm 0.5cm 4cm 2.5cm},clip]{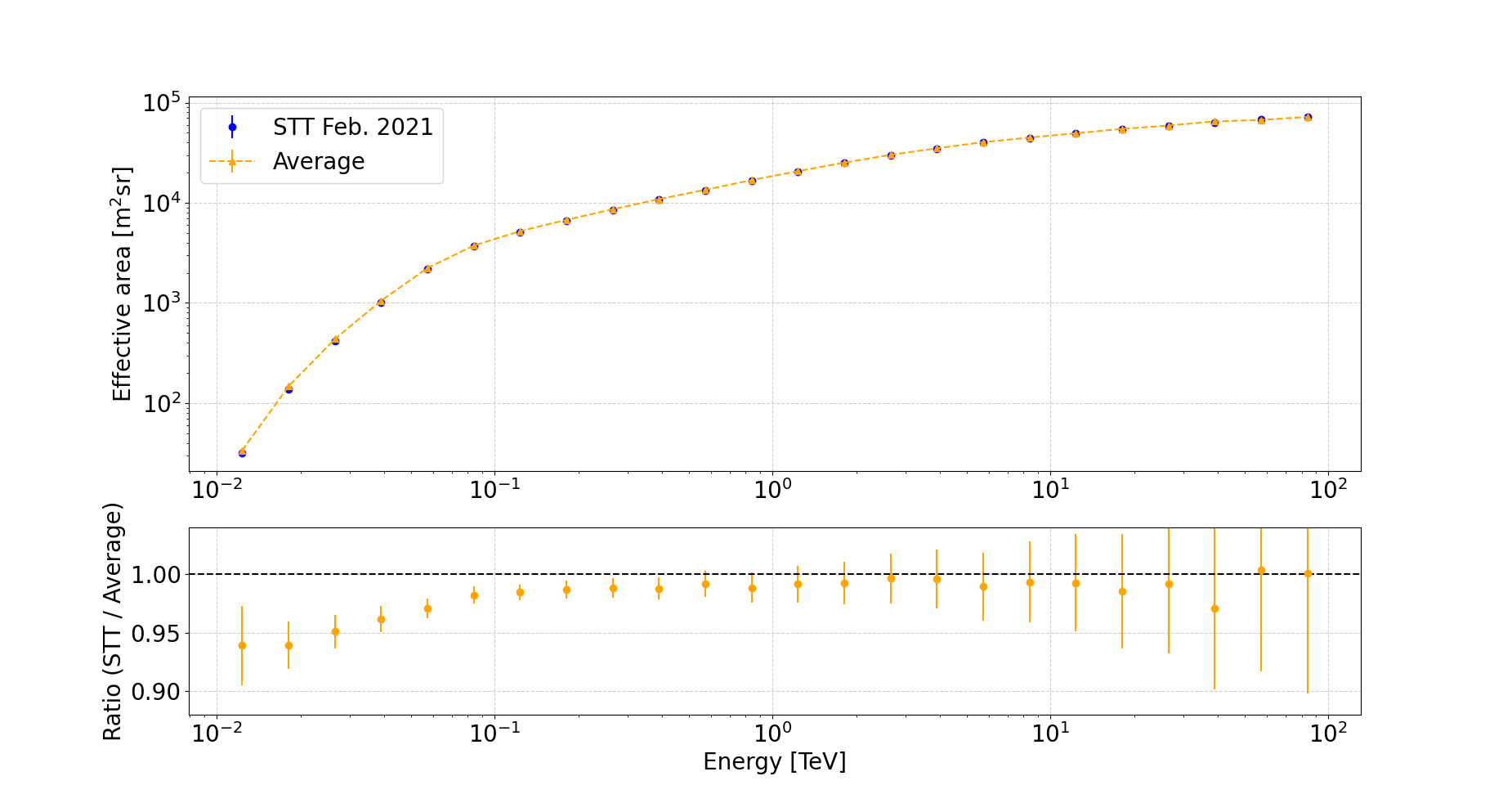}%
   \caption{Top panel: stereo trigger effective areas for the CTAO-North array under two different molecular extinction profiles: after incorporating the STT event of Feb. 2021, or the average ozone profile. Bottom panel: ratio of the two trigger effective areas.}%
   \label{fig:ea_north}
\end{figure}%

Figure~\ref{fig:ea} presents a comparison of the trigger effective areas (i.e., before reconstruction and analysis cuts) for the CTAO-South array under two different MEPs. A reduction in effective area is observed during the STT event, with the effect being significant at lower energies ($\leq$ 80~GeV), where it is expected to primarily impact the LSTs. At higher energies, the trigger effective area difference is within the uncertainty limits. Figure~\ref{fig:ea_north} shows the corresponding comparison for the CTAO-North. A similar energy-dependent reduction is observed, though slightly more pronounced. In this case, the decrease in effective area during the STT event reaches up to 5~\% for energies below 40~GeV.

\subsection{Systematic uncertainties affecting the study}
\label{sec:shortcuts}

This study is subject to systematic uncertainties arising from simplifications made in the analysis. The main simplifications are:
\begin{itemize}
    \item Neglecting the temperature dependence of absorption cross sections
    \item Omitting aerosol extinction profiles
    \item Uncertainties in the ozone profiles themselves     
\end{itemize}

\begin{figure}
  \centering
   \includegraphics[width=0.9\textwidth]{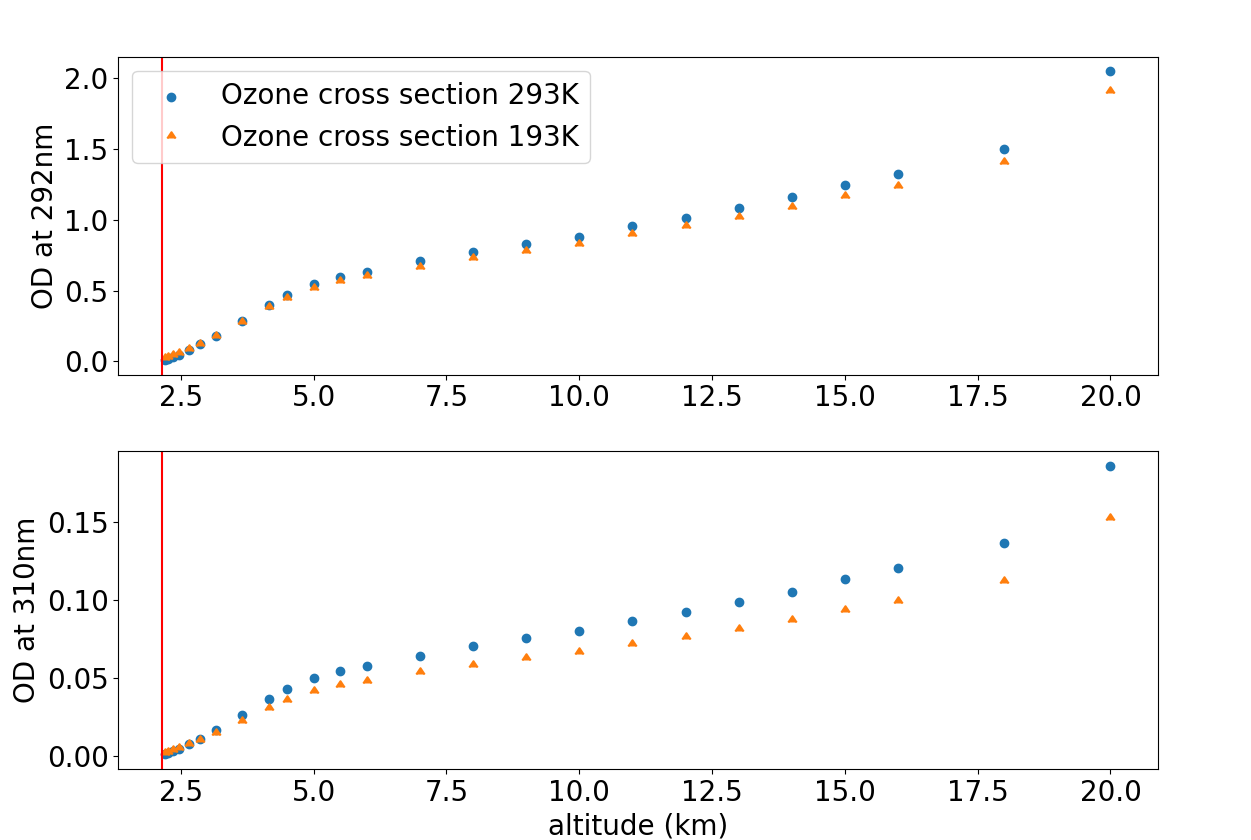}%
   \caption{Comparison of two ozone extinction optical depth profiles, calculated using absorption cross section at 293~K and 193~K. The top pannel shows the difference at a wavelength of 292 nm (Hartley band), while the bottom at a wavelength of 310 nm (Huggins band).}%
   \label{fig:temp}
\end{figure}

The temperature dependence of the ozone absorption cross section is primarily evident in the Huggins bands, where absorption is relatively weak. To quantify this effect, we generated two MAPs using the same ozone number density profile. In one case, the profile was multiplied by the absorption cross section at 293 K, while in the other, it was multiplied by the cross section at 193 K. These temperature values were chosen to approximately represent the limits of the atmospheric temperature range within the altitudes of interest. Figure~\ref{fig:temp} illustrates the comparison of the two optical depth profiles. As expected, the differences are more pronounced in the Huggins band, despite that fact that ozone absorption is not the strongest in this wavelength region. The two profiles were also examined with \textit{testeff}. The results of this analysis are presented in Table~\ref{tab:temp_dep}. The impact on Cherenkov light reaching the ground, if we consider only ozone-related extinction, can reach $\sim$8\% when examining the two extreme temperatures, 293~K vs. 193~K. Nevertheless, if we account for the telescope efficiency and the Rayleigh scattering, the effect due to the temperature dependence of the cross sections remains  $\leq0.3\%$. Given that we are comparing two extreme temperature cases relative to the expected atmospheric conditions, the choice to use only the ozone absorption cross section corresponding to a temperature of 293~K is justified. 
The ozone profiles have shown inter-model uncertainties below 10\% for the location of CTAO-North, but of the order of 20\% for CTAO-South~\cite{Arisio:2025}. The uncertainty of STT ozone concentrations is at least 20\%~\cite{Archibald:2020}.

\begin{table}
    \centering
    \begin{tabular}{|c|c|c|}
        \hline
        Extinction profile & Cross section at & LST telescope and atmosphere efficiency\\
        \hline
        ozone STT & 293~K & 
        0.1658
        \\
        \hline
        ozone STT & 193~K & 
        0.1664
        \\
        \hline
        MEP STT & 293~K & 
        0.1255
        \\
        \hline
        MEP STT & 193~K & 
        0.1258
        \\
        \hline
    \end{tabular}
    \caption{Effect of the temperature dependence of ozone absorption cross section on the Cherenkov light registered by LST. The first two rows show the difference if only ozone contributes to atmospheric extinction; while the last two show the results when the effect of Rayleigh scattering is accounted for as well. 
    }
    \label{tab:temp_dep}
\end{table}

The omission of the aerosol extinction profile is not expected to introduce a systematic bias in this work, as the analysis is based on relative comparisons. While the absence of aerosols decreases the absolute atmospheric transparency, it does not influence the relative variation in the recorded Cherenkov light associated with STT events.

\section{Discussion on calibration strategy}
\label{sec:disc} 

The CTAO requires a systematic uncertainty on the energy scale less than 10\% at 90\% confidence level. This budget can be further broken down~\cite{gaugSPIE2014} into contributions for the atmosphere~\cite{Gaug:2017Atmo}, the telescopes~\cite{mitchell2015,CTC:2019,GaugMuons:2019} and the analysis:
\begin{align}
\label{eq.energy}
\left( \frac{\delta E_\mathrm{scale}(E) }{E} = 10\% \right)^2 \quad &  \overset{!}{\lesssim}
  \left( \frac{\delta E_\mathrm{atmosphere}(E) }{E} = 8\% \right)^2  \quad + \nonumber\\[0.3cm]
& \quad + \quad \left( \frac{\delta E_\mathrm{telescopic~part}(E)}{E} = 5\% \right)^2  
  \quad + \quad \left( \frac{\delta E_\mathrm{analysis}(E)}{E} = 4\% \right)^2 \quad.
\end{align}
The atmosphere is the medium with which the primary gamma-rays interact and Cherenkov light is produced and propagates through. The development of the shower and the production of Cherenkov light depend on the molecular profile of the atmosphere while its propagation is subjected to various extinction processes, as mentioned in the introduction. One of them is the molecular absorption, and therefore its contribution is included in the overall atmospheric contribution. Preliminary estimates~\cite{Gaug:2017Atmo} indicated that controlling the contribution on the systematic uncertainty due to absorbing molecules at the $\sim$1\% level is sufficient to meet the requirement on the atmospheric contribution to the overall uncertainty budget for the energy scale. An analysis of the recorded image intensities during the selected STT events showed that the images are dimmer by (2--3)~\% for intensities below 1000 photoelectrons and by (1--2)~\% for intensities within the range of 1000 to \num{10000} photoelectrons. 

Regarding the effective area, the CTAO requirements restrain the systematic uncertainty budget related to atmospheric effects to less than (6--7)\% in the energy range from 100~GeV to 100~TeV. The observed impact of ozone variation on the trigger effective area was found to be energy-dependent, especially for shallow STT events, as expected, and of $\lesssim 1$\% for the energy range of the requirement.  At lower energies (particularly, below 40 GeV), the point-to-point variation in effective area is, however, larger and approximates 5\%. Therefore, studies focusing on this energy range could benefit from the application of an explicit ozone calibration.

Given that shallow STT events occur frequently—particularly from December to May for the Northern array and from June to January for the Southern array—monitoring such events is recommended. Since these events tend to produce energy-dependent effects, implementing ozone-specific calibration procedures could help mitigate biases in the reconstruction of low-energy showers.

\section{Conclusions}

This study examines the impact of atmospheric molecular absorption processes on the performance of the CTAO. The influence of these processes depends on the state of the atmosphere, which can vary on different timescales. This variability is more pronounced for non-well-mixed gases, which can fluctuate significantly from night to night.

As part of this study, we identified publicly available atmospheric datasets and developed software tools to process these data to generate Molecular Absorption Profiles (MAPs) and Molecular Extinction Profiles (MEPs). Characteristic atmospheric state profiles were incorporated into both fast and full simulation studies to assess their effects on the amount of Cherenkov light reaching the ground, image intensities in the telescope cameras, and the trigger effective areas. These studies contribute to refining the systematic uncertainty budget of CTAO and provide guidance for an improvement of CTAO's atmospheric calibration strategy for absorbing molecules.

Currently, there is no formal requirement to account for ozone or other absorbing molecules through calibration during CTAO operations. However, our findings indicate that ozone variability -- particularly during STT events -- can have a non-negligible impact on CTAO performance. While deep STT events are relatively rare, shallow events occur frequently during specific periods of the year at both CTAO sites. These events can lead to reductions of approximately 3~\% in image intensity and up to 5~\% in trigger effective area in the low-energy range.

As a result, regular monitoring of the ozone mixing ratio profile is recommended. Depending on the scientific objectives, the production of tailored ozone profiles in case an STT event happened could improve the accuracy of the Instrument Response Functions, especially at low energies. The tools developed in this work could be integrated into the calibration pipeline of the Data Processing and Preservation System (DPPS) of CTAO~\cite{dpps}, should a requirement for ozone calibration be established. In contrast, the impact of nitrogen oxide variability on CTAO performance was found to be negligible.

\bibliographystyle{JHEP}
\bibliography{biblio.bib}

\end{document}